\documentclass[aps]{revtex4-2}
\usepackage{graphicx}
\usepackage{color}
\def\bc{\begin{center}}
\def\ec{\end{center}}

\def\beq{\begin{equation}}
\def\eeq{\end{equation}}

\def\d{\downarrow}
\def\u{\uparrow}

\begin{document}

\title{Probing many-body systems near spectral degeneracies}

\author{K. Ziegler}
\affiliation{
Institut f\"ur Physik, Universit\"at Augsburg, D-86135 Augsburg, Germany}
\date{\today}

\begin{abstract}
The diagonal elements of the time correlation matrix are used to probe closed quantum systems that are measured
at random times. This enables us to extract two distinct parts of the quantum evolution, a recurrent part and an exponentially 
decaying part. This separation is strongly affected when spectral degeneracies occur, for instance, in the presence of
spontaneous symmetry breaking. Moreover, the slowest decay rate is determined by the smallest energy level spacing, and 
this decay rate diverges at the spectral degeneracies.
Probing the quantum evolution with the diagonal elements of the time correlation matrix is discussed as a general concept
and tested in the case of a bosonic Josephson junction. It reveals for the latter characteristic properties at the transition to
Hilbert-space localization. 
\end{abstract}

\maketitle

\section{Introduction}

Symmetries play a central role in classical as well as in quantum many-body systems. They determine the macroscopic
behavior of these systems. Moreover, symmetries of macroscopic states reflect symmetries and spontaneous symmetry
breaking of the underlying system. 
For instance, the groundstates of the Ising model with ferromagnetic nearest-neighbor spin-spin coupling are
$|\u,\dots,\u\rangle$ and $|\d,\dots,\d\rangle$. This two-fold degeneracy is a consequence of the symmetry
of the Hamiltonian. Then we experience spontaneous symmetry breaking when we add an arbitrarily small perturbation
(e.g., a small magnetic field or special boundary conditions) that breaks the symmetry. This leads to a lifting of
the degeneracy of the two quantum states. The fact that already an arbitrarily small symmetry breaking term
can lift the macroscopic degeneracy indicates a complex dynamics near this symmetry and the associated phase transition. 
In more general terms, it is important to understand the evolution of many-body systems near a symmetry.

In the following we will consider the unitary evolution of closed quantum many-body systems. It is based on the idea
that the extraction of information about the quantum system in an experiment is limited. In other words, not all properties 
or degrees of freedom of the quantum model are accessible by the experiment. Typical exceptions are the return and
transition probabilities for quantum states. Moreover,
quantum systems have a complex dynamics. Even for a few particles the evolution can be quite erratic similar to a classical 
random walk.
Such a behavior suggests a statistical approach to extract generic information about the quantum evolution, using
averaged quantities. A statistical approach is also 
supported by the fact that large sets of experimental data are available whose properties can be treated statistically. This idea is not new and found a very successful realization in the Random Matrix Theory (RMT).
It has been applied to many physical systems, such as nuclei, atoms and mesoscopic systems
\cite{10.2307/1970079,PhysRev.104.483,doi:10.1063/1.1703773,1966NucPh..78R.696L,
10.2307/2027409,mehta2004random,RevModPhys.69.731}.
The motivation for the RMT is that there is no way of knowing the Hamiltonian of even a relatively small many-body 
quantum system, such as an atomic nucleus. On the other hand, the spectra of these systems, complex though, have some 
characteristic features, such as level repulsion. Thus, instead of guessing a specific Hamiltonian, a random ensemble of
Hamiltonians is chosen, which describes the generic features of a class of quantum systems. 
The class is characterized by the invariance of the random ensemble with respect to symmetry transformations. These are 
typically orthogonal, unitary or symplectic transformations.
Another application of the RMT has been recently proposed for the description of random measurements.
It is based on Dyson's circular matrix ensemble 
\cite{doi:10.1063/1.1703773,doi:10.1063/1.1703774,doi:10.1063/1.1703775,mehta2004random}), 
which represents random unitary matrices and has been used as a tool to determine the trace of powers of the density matrix
and the related R\'enyi entropy
\cite{PhysRevLett.108.110503,PhysRevLett.120.050406,PhysRevA.97.023604,PhysRevB.100.134306,PhysRevX.9.031009}.

In contrast to these RMT approaches, we consider in the following a dynamical approach in which only the time of
a measurement is
random, whereas the energy levels $\{E_j\}$ of the Hamiltonian $H$ and the overlaps $\langle E_j|\Psi_0\rangle$, 
$\langle E_j|\Psi\rangle$ of the energy eigenstates $\{|E_j\rangle\}$ with a given initial state $|\Psi_0\rangle$ and 
a measured state $|\Psi\rangle$ are not random. This leads to the time correlation matrix (TCM) as the central tool
for the definition of the statistical model, instead of the random ensemble of Hamiltonians in the RMT.
We will employ this approach, which was previously described in Ref. \cite{Ziegler_2021}, to analyze the evolution
of the return and transition probabilities. In more concrete terms, for a given time $t_k$ we evaluate (in a calculation
or in a real experiment) the probability $p_k$ that the
system is in a certain state. Then we evaluate the probabilities $\{p_1, p_2, ...\}$ at different discrete and randomly
chosen times $\{t_1, t_2, ...\}$. This can be translated into practical observations, where it is assumed that each 
experiment is prepared 
in the same initial state and all measurements are performed for the same final state of the
evolution at different times. These experiments
provide an ensemble of probabilities $\{p_1, p_2, ...\}$ with the corresponding times $\{t_1, t_2, ...\}$. 

For given overlaps $\langle E_j|\Psi_0\rangle$, $\langle E_j|\Psi\rangle$ we can immediately predict some restrictions for
the evolution of the probability $p_k$ in the $N$--dimensional Hilbert space. When the overlaps vanish for some of the
eigenstates $|E_n\rangle$, the evolution cannot reach those states and the accessible Hilbert space is restricted to the states
$|E_j\rangle$ with $j\ne n$. This reduction of the Hilbert space can be interpreted as Hilbert-space localization
\cite{cohen16} or Hilbert-space fragmentation \cite{moudgalya2021hilbert}. 
This effect can be associated with spontaneous symmetry breaking, induced by the choice of the initial and measured
states.
In the case that the overlap to some states is not strictly zero but very small, the access to those
states may be negligible and can be ignored. This corresponds to a complex dynamical behavior and requires a 
careful analysis. It will be addressed briefly for the example of a bosonic Josephson junction in Sect. \ref{sect:bjj},
where the mirror symmetry of the junction is spontaneously broken.

This paper is organized as follows. After the definition of the TCM in Sect. \ref{sect:TCM} we focus on the properties 
of its diagonal elements (Sect. \ref{sect:diag}). Then the effect of spectral degeneracies 
on the diagonal TCM elements are discussed in Sect. \ref{sect:deg}.
In Sect. \ref{sect:bjj} we analyze the diagonal TCM elements in the specific example of a bosonic
Josephson junction.

\section{Time Correlation matrix}
\label{sect:TCM}

We consider the transition amplitude from $|\Psi_0\rangle$ to $|\Psi\rangle$
\beq
u_k=\langle\Psi|e^{-iHt_k}|\Psi_0\rangle
\ ,
\eeq
which is based on the unitary evolution with the Hamiltonian $H$ from the initial state $|\Psi_0\rangle$.
The probability to measure the state $|\psi\rangle$ at time $t_k$ is given by 
$|\langle\Psi|e^{-iHt_k}|\Psi_0\rangle|^2$. In other words, $p_k=|u_k|^2$ is the probability to find the 
quantum system in the state $|\Psi\rangle$ after the unitary evolution from the initial state $|\Psi_0\rangle$ 
over the time $t_k$. Since the evolution is defined by the Hamiltonian $H$, we consider its eigenstates 
$\{|E_j\rangle\}_{j=1,...,N}$ and its corresponding eigenvalues $\{E_j\}_{j=1,...,N}$ and
write the amplitude in spectral representation as
\beq
u_{k}=\sum_{j=1}^N \langle\Psi|E_j\rangle\langle E_j|\Psi_0\rangle e^{-iE_jt_k}
\equiv \sum_{j=1}^N q_j e^{-iE_jt_k}
\ .
\label{spectr_rep1a}
\eeq
Although the phases are not directly 
accessible experimentally, their effect is observable through the interference of quantum states. This enables us,
for instance, to relate the product of amplitudes at different times with probabilities of interfering amplitudes:
\beq
u_k^*u_{k'}+u_{k'}^*u_{k}
=\frac{1}{2}\left(|u_k+u_{k'}|^2 - |u_k-u_{k'}|^2\right)
\ ,\ \ 
i(u_k^*u_{k'}-u_{k'}^*u_{k})
=\frac{1}{2}\left(|u_k+iu_{k'}|^2 - |u_k-iu_{k'}|^2\right)
\ ,
\eeq
where the probabilities are observable in interferometric measurements. 
This relation suggests to consider the correlation of the amplitudes $u_k$, $u_{k'}$ at different times 
through the TCM $\langle u^*_k u_{k'}\rangle_\tau$, where the average $\langle ...\rangle_\tau$
is taken with respect to the distribution of times $\{t_k\}$ as a result of inaccurate clocks:
The time is measured by a clock in each laboratory, which counts time steps $\{ \tau_n\}$.
These clocks have a limited accuracy, such that the time steps vary randomly. 
This implies a sequence of measurements in each laboratory, where the clock indicates $k$ time steps corresponding 
to the total evolution time $t_k=\tau_1+\cdots+\tau_k$ for different values of $k$. 
Now we compare the
measurements of different laboratories. This provides a distribution of results for $u^*_k u_{k'}$ due to different
inaccurate clocks, where we assume that the fluctuations of the time steps $\{\tau_n\}$ are independently
and equally distributed. 
Then the TCM reads in spectral representation
\beq
\langle u_k^*u_{k'}\rangle_\tau=
\sum_{j,j'}q^*_jq_{j'}\langle e^{iE_j(\tau_1+\cdots+\tau_k)}e^{-iE_{j'}(\tau_1+\cdots+\tau_{k'})}\rangle_\tau
\eeq
\beq
=\sum_{j,j'}q^*_jq_{j'}\cases{
\langle e^{i(E_j-E_{j'})(\tau_1+\cdots+\tau_k)}\rangle_\tau\langle e^{-iE_{j'}(\tau_{k+1}+\cdots
+\tau_{k'})}\rangle_\tau & $k'>k$ \cr
\langle e^{i(E_j-E_{j'})(\tau_1+\cdots+\tau_{k'})}\rangle_\tau\langle e^{iE_{j'}(\tau_{k'+1}
+\cdots+\tau_{k})}\rangle_\tau & $k'<k$ \cr
\langle e^{i(E_j-E_{j'})(\tau_1+\cdots+\tau_{k})}\rangle_\tau & $k'=k$ \cr
}
\ .
\label{product_ua}
\eeq
Defining $\lambda_j=\langle e^{iE_j\tau}\rangle_\tau$ and $\lambda_{jj'}=\langle e^{i(E_j-E_{j'})\tau}\rangle_\tau$
the TCM elements become
\beq
\langle u_k^*u_{k'}\rangle_\tau=\sum_{j,j'}q^*_jq_{j'}\cases{
\lambda_{jj'}^k\lambda_{j'}^{k'-k}&  $k'\ge k$ \cr
\lambda_{jj'}^{k'}{\lambda^*_{j}}^{k-k'}& $k'<k$ \cr
}
\ .
\label{u_product}
\eeq
The TCM decays exponentially with $|k-k'|$, provided $|\lambda_j|<1$. 
For fixed $|k-k'|$ the TCM is constant for the diagonal elements $\lambda_{jj}=1$, 
though. This reflects the fact that a unitary evolution between the same energy eigenstates
gives just a phase factor $e^{-iE_j\tau}$ (cf. Eq. (\ref{spectr_rep1a})). On the other hand, 
these phase factors lead to a decay of different energy states due to interference effects after
the time average. 

\subsection{Diagonal elements of the TCM}
\label{sect:diag}

The diagonal TCM element $\langle |u_k|^2\rangle_\tau$ is the probability to measure the state $|\Psi\rangle$
at time $t_k$. Before time averaging, the expression $|u_k|^2$ is a diagonal element of the density matrix 
$\rho(t_k)$ with respect to the state $|\Psi\rangle$. The trace of $|u_k|^2$ with respect to all states $|\Psi\rangle$ 
of the underlying Hilbert space is the spectral form factor, often used for the characterization of many-body quantum chaos
\cite{PhysRevX.8.021062,PhysRevX.8.041019}. We only mention this but will not study it here.

According to Eq. (\ref{u_product}) the average transition probability $\langle |u_k|^2\rangle_\tau$ reads
\beq
\langle |u_k|^2\rangle_\tau
=P_N+\sum_{j,j'=1;j'\ne j}^Nq^*_jq_{j'}\lambda_{jj'}^k
\ ,\ \ 
P_N=\sum_{j=1}^N|q_j|^2
\ .
\label{diag_elements00}
\eeq
The term $P_N$ describes the recurrent behavior, which does not depend on time. It is the asymptotic
transition probability for $k\to\infty$
\beq
P_N=\lim_{k\to\infty}\langle |u_k|^2\rangle_\tau
\ ,
\eeq
provided that the energy levels are not degenerated. The case of degenerate energy levels is discussed in the next section.
The second term in Eq. (\ref{diag_elements00}) decays exponentially with time due to $|\lambda_{jj'}|<1$, and
only this term describes a change of the transition probability during the evolution of the quantum system.
This result provides a separation of the diagonal elements of the TCM into a static recurrent term $P_N$ and a dynamic
term that decays quickly.

The recurrent term $P_N$ stores important information regarding the
properties of the quantum system. Since $|q_j|^2=|\langle\Psi|E_j\rangle|^2|\langle\Psi_0| E_j\rangle|^2$
is a product of the overlaps between the energy eigenstate $|E_j\rangle$ with the initial state
and with the measured state, it provides a measure of how much this energy eigenstate contributes to the
transition $|\Psi_0\rangle\to |\Psi\rangle$ during the unitary evolution. For instance, the asymptotic behavior
of the return probability to the initial state $|\Psi_0\rangle\to |\Psi_0\rangle$ with the dimensionality $N$ of
the underlying Hilbert space describes Anderson localization when $\lim_{N\to\infty}P_N>0$ and the
absence of Anderson localization when $\lim_{N\to\infty}P_N= 0$ \cite{cohen16}. This can be understood
by noting that the normalization of quantum states implies $\sum_{j=1}^N|\langle\Psi_0| E_j\rangle|^2=1$ 
and that for a localized state only a few energy eigenstates have a nonzero overlap with $|\Psi_0\rangle$. For a delocalized
state, on the other hand, the overlap is nonzero for a large number of energy eigenstates which is
of the order of $N$. An extreme case is given when these overlaps are equal. Then we have 
$|\langle\Psi_0| E_j\rangle|^2=1/N$ due to the normalization, which implies $P_N=1/N$.
Anderson localization is associated with a random Hamiltonian \cite{PhysRev.109.1492}. According to the above
described picture, we can also consider Hilbert-space localization for a deterministic Hamiltonian, which depends strongly
on the initial state. For an energy eigenstate, the system will always
remain in the latter under unitary evolution. More general, is the initial state a superposition of $m$ energy eigenstates,
the system will always remain inside the $m$--dimensional Hilbert space, spanned by these energy eigenstates. In the
case, where the initial state is eigenstate of $H_0$ of the Hamiltonian $H=H_0+\eta H_1$ and 
$\eta$ is a small parameter, $\eta H_1$ is a small perturbation. 
In that case it is possible that this perturbation provides an exponentially decaying evolution away the initial state. 
This would be considered as exponential Hilbert-space localization.

In the subsequent discussion we will focus on the diagonal elements of the TCM, since
the off-diagonal TCM elements decay exponentially with $|k-k'|$ according to Eq. (\ref{u_product}).

\subsection{Effect of spectral degeneracies}
\label{sect:deg}

Assuming that there is a spectral degeneracy $E_1=E_2$, we have $\lambda_{12}=\lambda_{21}=1$, and the diagonal 
TCM elements in Eq. (\ref{diag_elements00}) read in this case
\beq
\langle |u_k|^2\rangle_\tau
=\sum_{j,j'=1}^Nq^*_jq_{j'}\lambda_{jj'}^k
=P_N+q_1^*q_2+q_2^*q_1+\sum_{j,j'=1;j'\ne j;(j,j')\ne(1,2),(2,1)}^Nq^*_jq_{j'}\lambda_{jj'}^k
\ ,
\label{diag_elements02}
\eeq
such that the recurrent part of the transition probability becomes 
\beq
\lim_{k\to\infty}\langle |u_k|^2\rangle_\tau
=P_N+q_1^*q_2+q_2^*q_1=|q_1+q_2|^2+\sum_{j=3}^N|q_j|^2
\ . 
\eeq
Thus, the effect of a spectral degeneracy is a change of the recurrent and the decaying behavior, where the
recurrent term changes by $|q_1+q_2|^2-|q_1|^2-|q_2|^2$.  
This means that the diagonal elements of the TCM are very sensitive in terms of spectral degeneracies.

After applying a discrete Fourier transformation to the decaying part of $\langle |u_k|^2\rangle_\tau$ we obtain the function
\beq
{\tilde U}_d(e^{i\omega})=
\sum_{k\ge1}e^{i\omega k}\sum_{j,j'=1;k'\ne k;(j,j')\ne(1,2),(2,1)}^Nq^*_jq_{j'}\lambda_{jj'}^k
=\sum_{j,j'=1;j'\ne j;(j,j')\ne(1,2),(2,1)}^Nq^*_jq_{j'}\frac{\lambda_{jj'}}{e^{-i\omega}-\lambda_{jj'}}
\ ,
\eeq
which is a function of $\omega$ on the interval $[0,2\pi)$. In other words,  ${\tilde U}_d(z)$ is a
sum of poles inside the unit circle due to $|\lambda_{jj'}|<1$. The poles $\lambda_{nn'}$ and $\lambda_{n'n}$
approach the unit circle when we are getting closer to a degeneracy of $E_n$ and $E_{n'}$ This should be  visible in 
${\tilde U}_d(e^{i\omega})$. The corresponding decay time 
$T_d=-1/\log|\lambda_{nn'}|$ diverges due to $|\lambda_{nn'}|\sim1$.
Therefore, the decay time $T_d$ is a measure for the distance from a spectral degeneracy; it diverges when we
approach the degeneracy. In general, we can define
\beq
T_m=\max_{j,j'=1,\ldots, N} -\frac{1}{\log|\lambda_{jj'}|}
\eeq
as the largest decay time as a measure of level degeneracy.

\section{Example: bosonic Josephson junction}
\label{sect:bjj}

In this section we will study the diagonal TCM elements of a bosonic Josephson junction (BJJ) with $N$ bosons
as a closed quantum system. The motivation for choosing this example is at least threefold: The model is (i) simple 
enough but not trivial with interesting features based on tunneling and boson-boson interaction, (ii) it can be solved 
exactly and (iii) it has been realized experimentally \cite{PhysRevA.80.053613,PhysRevA.86.023615}
with applications to commercial quantum computers \cite{PhysRevLett.111.080502}. The BJJ consists of
two identical wells filled with interacting bosons and a tunneling junction between them. More formally, it is defined by 
the Bose-Hubbard Hamiltonian \cite{Gati_2007}
\beq
H=
-\frac{J}{2}(a_l^\dagger a_r + a_r^\dagger a_l)+ U(n_l^2+n_r^2) , \ \ \ 
n_{l,r}=a_{l,r}^\dagger a_{l,r}
\ ,
\label{ham00}
\eeq
where $a^\dagger_{l,r}$ ($a_{l,r}$) are the creation (annihilation) operators in the left and right well, 
respectively. 
The first term of $H$ describes tunneling of atoms between the wells, and for $U>0$ the second 
term represents a repulsive particle-particle interaction that favors energetically
a symmetric distribution of bosons in the double well. Without tunneling (i.e., for $J=0$) there are two-fold 
degenerate energy levels $E_k=U[(N-k)^2+k^2]/2$ with eigenstates that are superpositions of the product 
Fock state $|k,N-k\rangle$ ($\equiv |k\rangle\otimes|N-k\rangle$) and its mirror image $|N-k,k\rangle$.
This two-fold degeneracy is similar to the two-fold degeneracy of the Ising model, mentioned in the Introduction.
Thus, tunneling between the wells plays the role of the symmetry-breaking term. In contrast to the macroscopic
Ising model though, an arbitrarily small tunneling term may not be sufficient to cause symmetry breaking. Hence the 
following study is more related to an Ising model with a finite number of spins than to a macroscopic system. Nevertheless,
the sensitive dynamics due to tunneling between the two wells with degenerate energy levels will give us an insight
into the evolution near symmetry points.
Another difference between the Ising model and the BJJ is that without tunneling in the latter 
all energy levels
are two-fold degenerate. Therefore, the initial state can be prepared in any of these energy 
levels to follow the evolution due to tunneling in the vicinity of degenerate levels. 
This is important because the degenerate groundstate may not be reached due to the energy
conservation in the closed quantum system.

For the following we will use $|\Psi_0\rangle=|0,N\rangle$ 
as the initial state and $|\Psi\rangle=|N,0\rangle$ as the measured state. Then we define the return probability (RP) 
and the transition probability (TP) as
\beq
|u_{r,k}|^2=|\langle0,N|e^{-iHt_k}|0,N\rangle|^2
\ ,\ \ \
|u_{t,k}|^2=|\langle N,0|e^{-iHt_k}|0,N\rangle|^2
\ .
\eeq
Since both states $|\Psi_0\rangle$, $|\Psi\rangle$ are eigenstates of $H$ in the absence of tunneling ($J=0$), we get
\beq
|u_{r,k}|^2=1
\ ,\ \ \ 
|u_{t,k}|^2=0
\eeq
for any $k$ due to orthogonality. This reflects the fact that this pair of states breaks the mirror symmetry of the double well.
The opposite extreme is the BJJ without boson-boson interaction ($U=0$), which is more complex and
will be discussed in the next subsection. As we will see, this case can be described
by simple functions for $|u_{r,k}|^2$, $|u_{t,k}|^2$. For the interplay of tunneling and boson-boson interaction 
($J,U\ne0$) the behavior is more complex and we will rely on the time averaged expressions 
$\langle|u_{r,k}|^2\rangle_\tau$,
$\langle|u_{t,k}|^2\rangle_\tau$ with $\tau_k={\bar\tau}+\tau'_k$ and an exponential distribution for $\tau'_k$:
\beq
\langle ...\rangle_\tau=\int_0^\infty ... \prod_{n\ge1}e^{-\tau'_n}d\tau'_n
\ ,
\label{exp_dist} 
\eeq
where the time is measured in units of $\hbar/J$.

\subsection{Non-interacting bosons}

For $U=0$ the spectrum of $H$ consists of equidistant energy levels $E_j=-J(N/2-j)$ ($j=0,1,...,N$)  and eigenstates
\[
|E_j\rangle=\frac{2^{-N/2}}{\sqrt{j!(N-j)!}}(a_l^\dagger+a_r^\dagger)^j(a_l^\dagger-a_r^\dagger)^{N-j}|0,0\rangle
\ .
\]
Then the RP and the TP at time $t_k$ for $N$ bosons read
\beq
|u_{r,k}|^2=|\langle 0,N|e^{-iHt_k}|0,N\rangle|^2
=|\cos^N(J t_k/2)|^2
\ ,\ \ \
|u_{t,k}|^2=|\langle N,0|e^{-iHt_k}|0,N\rangle|^2
=|\sin^N(J t_k/2)|^2
\ .
\eeq
For the average TP we get with $t_k=k{\bar\tau}+\tau'_1+\cdots + \tau'_k$ and the exponential distribution
of Eq. (\ref{exp_dist})
\[
\langle|u_{t,k}|^2\rangle_\tau
=2^{-2N}\sum_{l,l'=0}^N  {{N}\choose{l}}{{N}\choose{l'}}(-1)^{l+l'} 
\left[\frac{e^{-iJ{\bar\tau}(l-l')}}{1-iJ(l-l')}\right]^k
\]
\beq
=2^{-2N}\sum_{l=0}^N  {{N}\choose{l}}^2
+2^{-2N}\sum_{l,l'=0;l'\ne l}^N  {{N}\choose{l}}{{N}\choose{l'}}(-1)^{l+l'} 
\left[\frac{e^{-iJ{\bar\tau}(l-l')}}{1-iJ(l-l')}\right]^k
\ ,
\eeq
while for the average RP is the same without the factor $(-1)^{l+l'}$. From these results we get for the 
asymptotic value at $k\sim\infty$ for both probabilities
\beq
P_N\sim\langle|u_{r,k}|^2\rangle_\tau\sim\langle|u_{t,k}|^2\rangle_\tau\sim 2^{-2N}\sum_{l=0}^N  {{N}\choose{l}}^2
\ ,
\eeq
which reflects the mirror symmetry of the BJJ.

\begin{figure}[t]
\begin{center}
\includegraphics[width=9cm,height=6cm]{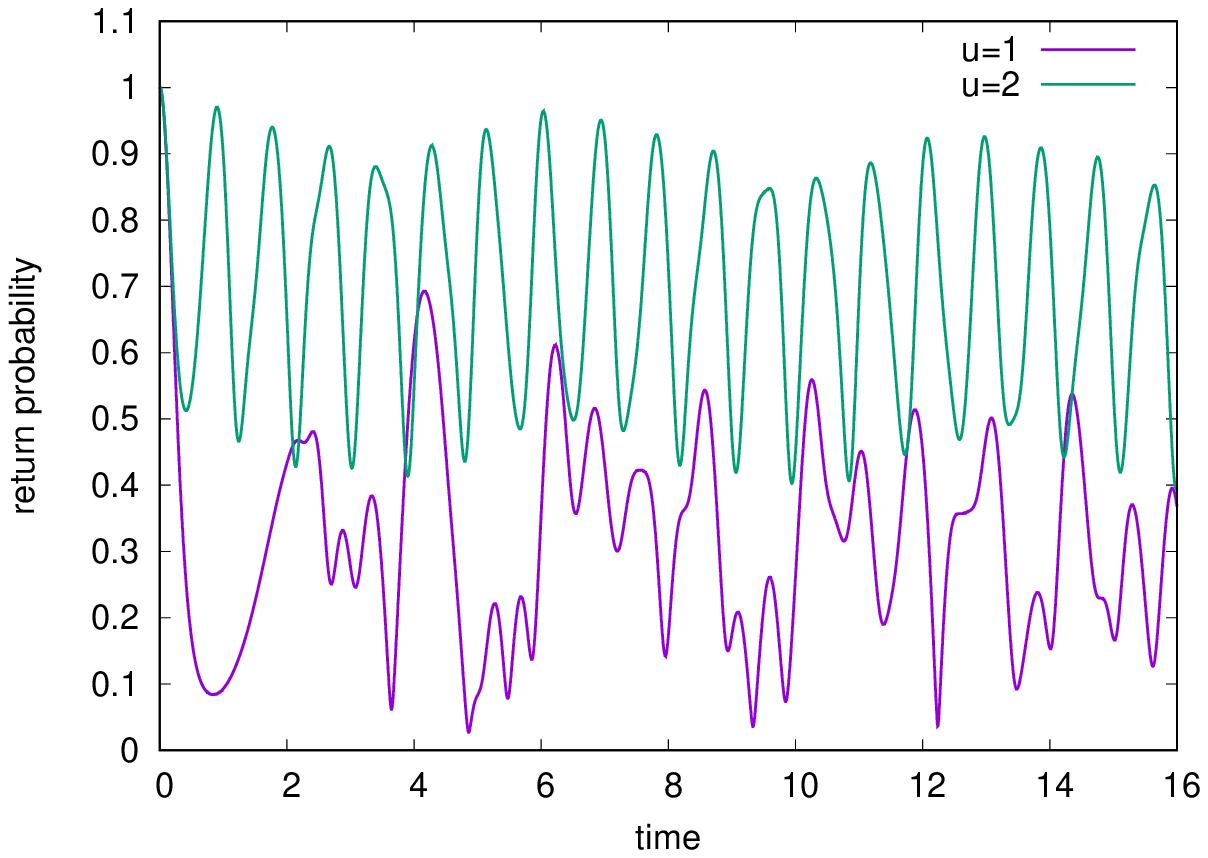}
\includegraphics[width=9cm,height=6cm]{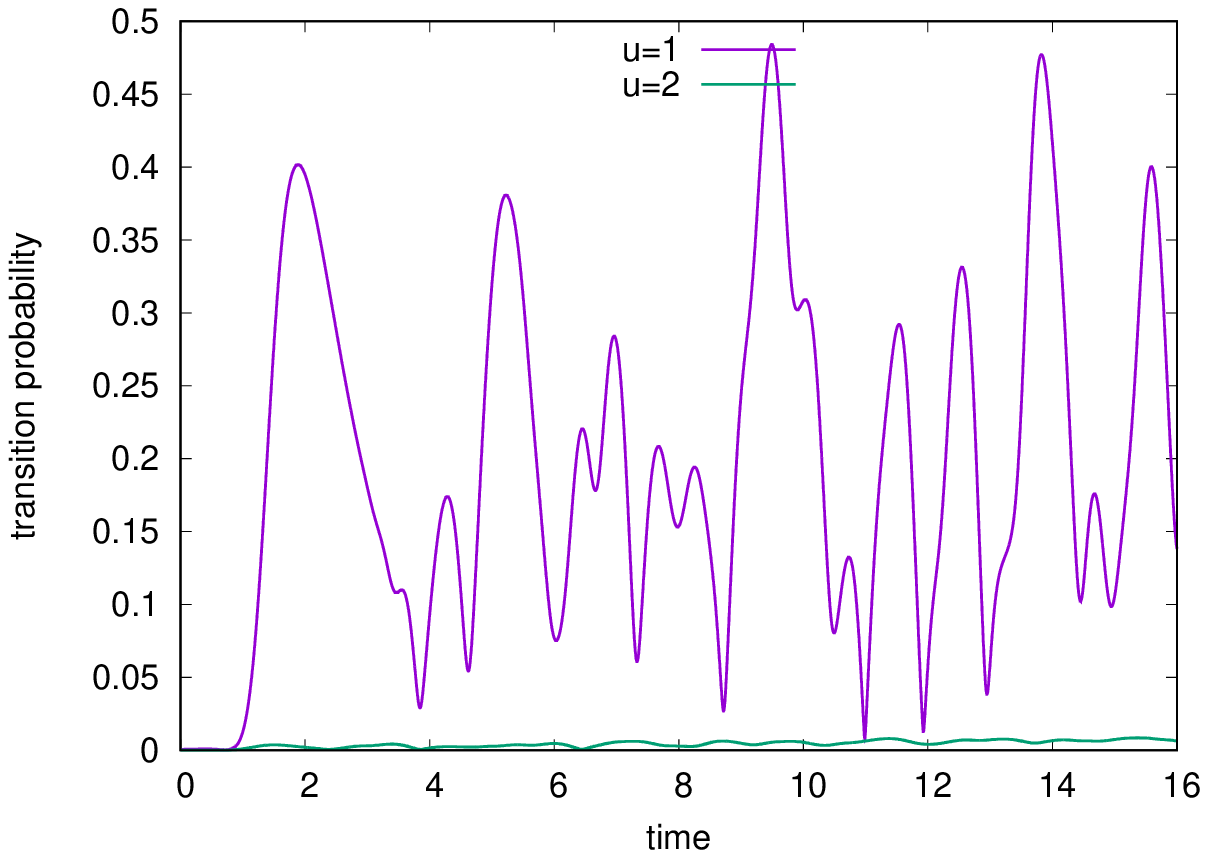}
\caption{
Signatures of a qualitative change of the evolution of a bosonic Josephson junction upon an increasing interaction
strength $u$. The plots represent the dynamics of 20 bosons for $u=1,2$, where the top panel gives the return probability 
$|u_{r,k}|^2$ and the bottom panel the transition probability $|u_{t,k}|^2$. 
}
\label{fig:dynamics1}
\end{center}
\end{figure}

\subsection{Interacting bosons}

Exact solutions exist for this model also for $J,U\ne 0$ but in contrast to the non-interacting case they are complex and
difficult to present in general. For instance, the resolvent is a meromorphic function with polynomials of order $N$
and $N+1$ \cite{cohen16,Ziegler_2011}. Therefore, we only plot the results for the RP and the TP and their averaged counterparts here.
For transparency, we choose for all subsequent plots $N=20$ bosons.

First, the evolution of the RP and the TP are presented in Fig. \ref{fig:dynamics1} for two values 
$u=1, 2$ of the interaction parameter $u=NU/J$. This clearly reveals that the RP
dominates over the TP for increasing $u$, as we expect from the results of the two limits
$J=0$ and $U=0$. It is interesting to note that in a mean-field (classical) approximation of the BJJ there is a sharp
phase transition in terms of the interaction parameter, where the mean-field TP is completely suppressed when
$u\ge u_c=2$ \cite{PhysRevA.55.4318}. The strong interaction phase is also called the self-trapping phase. The analogue
of the latter in the quantum BJJ is Hilbert-space localization, reflected by the scaling behavior of the inverse participation
ratio \cite{cohen16}. This also indicates the existence of a critical $u_c$.  

\begin{figure}[t]
\begin{center}
\includegraphics[width=9cm,height=6cm]{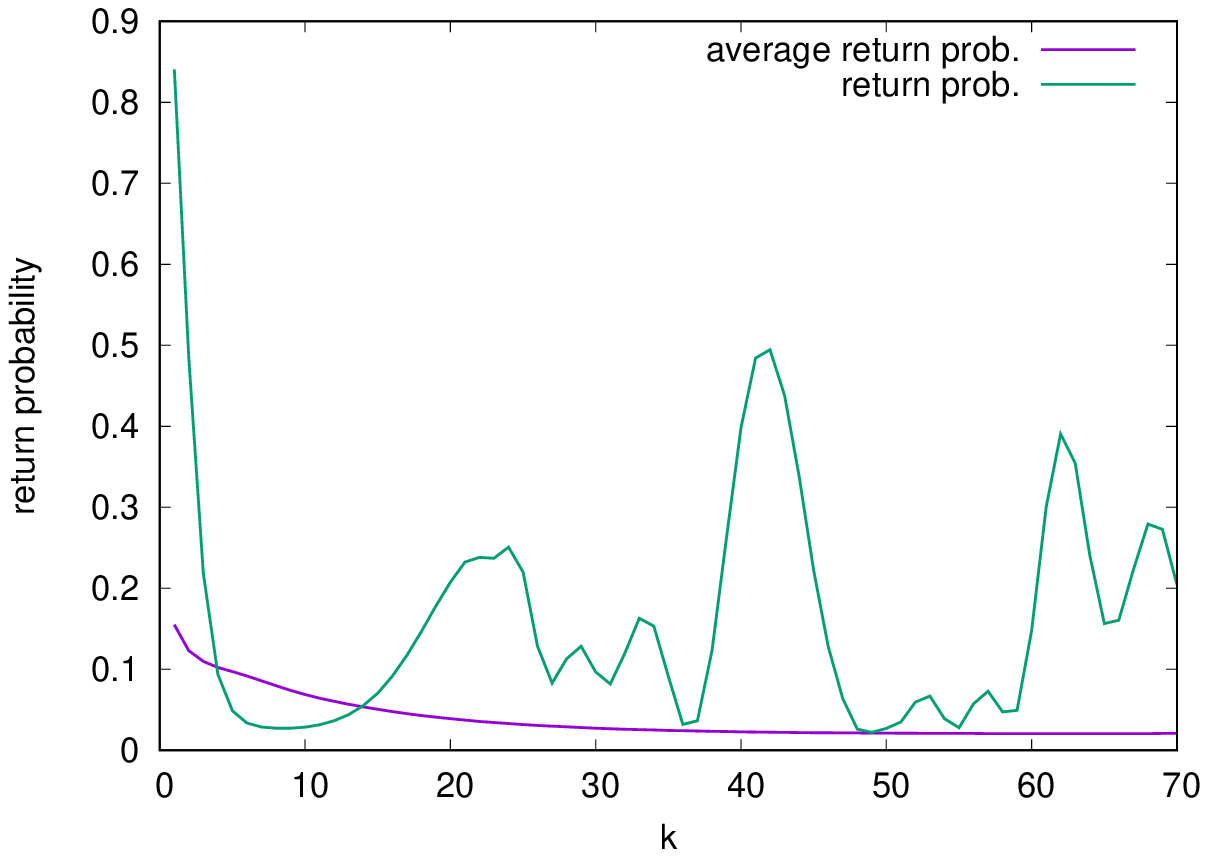}
\includegraphics[width=9cm,height=6cm]{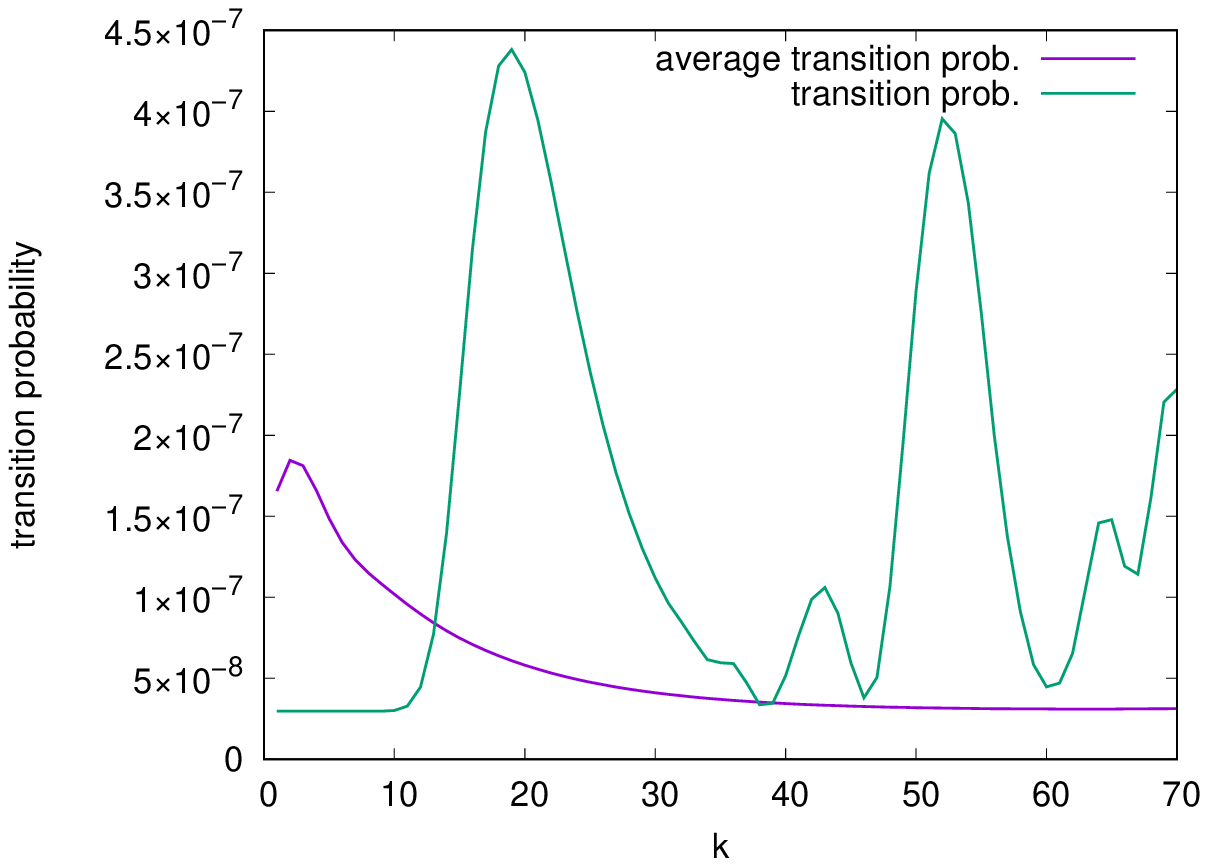}
\caption{
Comparison of the return probability $|u_{r,k}|^2$ and the average return probability $\langle|u_{r,k}|^2\rangle_\tau$
(top panel) and of the corresponding transition probabilities (bottom panel). The average was taken with respect to the
exponential distribution of Eq. (\ref{exp_dist}). The interaction parameter is $u=1$ and ${\bar\tau}=1/10$.
}
\label{fig:dynamics2}
\end{center}
\end{figure}

In Fig. \ref{fig:dynamics2} the effect of time averaging on $|u_{r,k}|^2$ and $|u_{t,k}|^2$ for $u=1$ is visualized.
It reflects the smoothing of the strongly fluctuating dynamics with a recurrent and a decaying contribution according to
Eqs. (\ref{diag_elements00}), (\ref{diag_elements02}). It is obvious that the separation of the recurrent and the decaying
behavior is not accessible without time averaging.

\begin{figure}[t]
\begin{center}
\includegraphics[width=9cm,height=6cm]{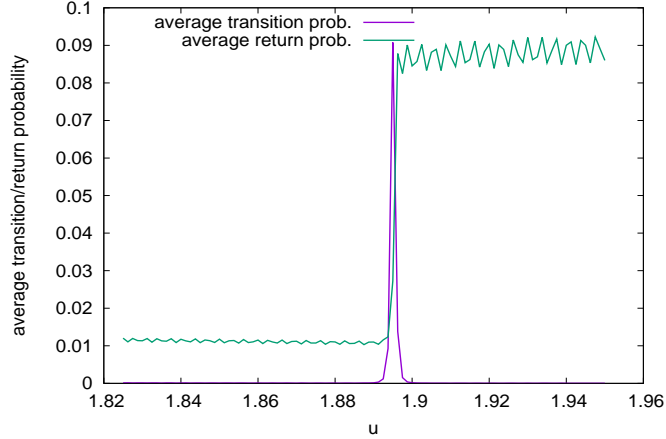}
\caption{
The critical regime of the Hilbert-space localization with $u_c\approx 1.89$ is visualized with 
$\langle|u_{r,k}|^2\rangle_\tau$  and $\langle|u_{t,k}|^2\rangle_\tau$ at $k=70$.
}
\label{fig:transition}
\end{center}
\end{figure}

The existence of a critical interaction strength $u_c\approx 1.89...$ is demonstrated in Fig. \ref{fig:transition},
where the $\langle|u_{r,k}|^2\rangle_\tau$ jumps up upon increasing $u$ at $u_c$. Moreover, $\langle|u_{t,k}|^2\rangle_\tau$ 
develops a characteristic peak at $u_c$.
This behavior reflects the appearance of nearly degenerate energy levels, as described in Sect. \ref{sect:deg}.

\begin{figure}[t]
\begin{center}
\includegraphics[width=9cm,height=6cm]{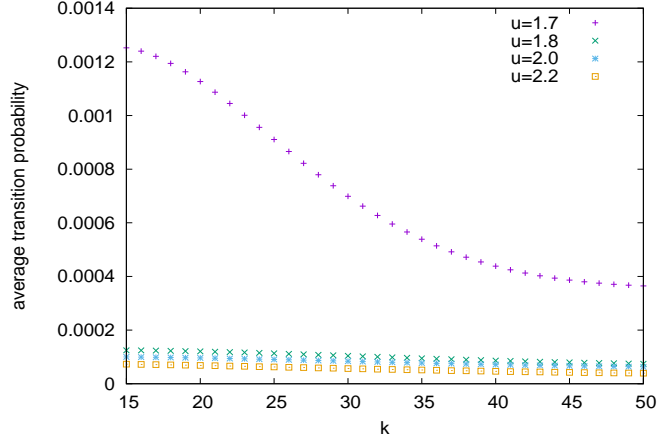}
\caption{
Decay of $\langle|u_{t,k}|^2\rangle_\tau$ for different interaction parameters $u=1.7,\ldots ,2.2$.
}
\label{fig:decay}
\end{center}
\end{figure}

Finally, in Fig. \ref{fig:decay} the change of the time scales for the decay of the average TP $\langle|u_{t,k}|^2\rangle_\tau$ 
is visualized for $u=1.7,...,2.2$. The decay is reduced by an increasing interaction strength $u$. This reflects the fact 
that the splitting of the energy levels is reduced by the interaction, as we would have expected.

\section{Discussion and conclusions}

Our analysis of the quantum unitary evolution is strictly focused on the result of a single measurement in each of many
identical experiments, which are subject of a unitary evolution. Averaging with respect to the statistical outcome due 
to measurements at randomly distributed times leads to the TCM. We have focused on the diagonal TCM elements to
study the evolution of the quantum system. The analysis of the off-diagonal TCM elements was the subject of a previous
work \cite{Ziegler_2021}. Similar to the off-diagonal TCM elements, the diagonal TCM elements reveal a separation of 
the evolution into a static recurrent part and a dynamic decaying part. We have found that the decay rate of the latter
is related to the spacing between energy levels, which diverges when the spacing vanishes near a degeneracy.
Thus, the decay rate is a quantity that can be used to detect symmetry changes or the appearance of spontaneous 
symmetry breaking. This has been observed in the example of the BJJ: In this model the energy levels are two-fold
degenerate in the limit $u\to\infty$. This is reflected in Fig. \ref{fig:decay}, where the decay decreases with increasing $u$. 

Another interesting aspect of the BJJ is the transition to Hilbert-space localization \cite{cohen16}. This transition was
also detected with the helps of the average RP and average TP in Fig. \ref{fig:transition}, where the average RP
experience a jump to a higher value for $u>u_c$. On the other hand, the average TP has only a sharp peak near $u_c$
but has the same value away from $u_c$.

We can conclude that time averaging over an ensemble of measurements is crucial for extracting the (static) recurrent
behavior and the (dynamic) decaying behavior. This can be formulated in terms of the TCM.
The separation of static and dynamic behavior is generic for the unitary quantum evolution. Then the
TCM provides a tool to 
analyze properties of the quantum system, which are associated with spectral degeneracy in the case of phase transitions.
It can be applied to theoretical calculations as well as to experimental data, collected from many experiments at
identical quantum systems. 
We have demonstrated in the case of the BJJ that the TCM approach delivers interesting generic information.
The BJJ can be considered as a building block of the Bose-Hubbard model on a lattice. Therefore, the TCM approach should 
be applicable to more complex quantum models, including bosonic and fermionic Hubbard models as well as
quantum spin systems. 

\acknowledgments{I am grateful to Eli Barkai for interesting discussions regarding the Hilbert-space
fragmentation. This work was supported by the Julian Schwinger Foundation.}

\bibliography{ref.bib}

\end{document}